# Dielectric Metasurface for Generating Broadband Millimeter Wave Orbital Angular Momentum Beams

Fan Bi[1,2], Zhongling Ba[2], Yunting Li[2], and Xiong Wang[2,†]
1 Shanghai Institute of Microsystem and Information Technology of the Chinese Academy of Sciences, Shanghai 200050, China
2 School of Information Science and Technology, ShanghaiTech University, Shanghai 201210, China
† To whom correspondence should be addressed; Email: wangxiong@shanghaitech.edu.cn

**Abstract:** Electromagnetic waves carrying orbital angular momentum (OAM) have found many applications. This work reports a broadband millimeter wave OAM generator applying a dielectric metasurface with high-refractive-index dielectric elements distributed on a low-refractive-index substrate. Proper adjustment of the dimensions of the dielectric elements can make a broadband OAM generator from 53 to 70 GHz. OAM generators with different mode numbers are designed and validated by simulation. Such structures can find potential applications in high-capacity communications and imaging.

## 1. Introduction

Electromagnetic waves are capable of bearing both spin angular momentum and orbital angular momentum (OAM) [1]. The former pertains to polarization and the latter is associated with a helical or twisted wavefront [1]. OAM beams have witnessed a tremendous amount of research activity in the past two decades owing to their interesting unconventional properties and exciting potential applications [2-4]. Theoretical investigation and exploration of diversified applications of OAM beams have been extensively reported first in the optical regime [5], where they are referred to optical vortex beams, and subsequently in lower frequency domain like terahertz [6], millimeter wave [7], microwave [8] and RF band [9]. Among various novel applications of OAM beams, the capability of offering a brand-new mechanism of data multiplexing for high-capacity communications is probably the most intriguing one and has fueled widespread interests [10].

An OAM beam is characterized with a helical wavefront described by an azimuthal phase dependence $e^{jm\varphi}$, where m is called OAM mode number and $\varphi$ is transverse azimuthal angle. A couple of approaches have been proposed to generate OAM beams with such azimuthal phase dependence in the microwave and millimeter wave range. The first way is using properly engineered structures like spiral phase plates [11] or flat phase plate [12]. The second technique makes use of circular waveguides [13]. The third method is based on metasurface [14], which forms a spiral phase shift profile on the metasurface. The fourth one takes advantage of a phased array antenna [15]. However, most of the reported OAM beam launchers are designed merely for a specified frequency point, which have very limited bandwidth and restrict some potential applications. For example, in communication applications broadband OAM generators can enable frequency-division multiplexing beside the OAM multiplexing, further boosting data capacity and spectral efficiency. Likewise, broadband OAM beams can help to enhance the resolution in related imaging applications.

In this work, generation of broadband millimeter wave OAM beams operating from 53 to 70 GHz, leading to a fractional bandwidth of 27.6%, based on a reflection-type dielectric metasurface is proposed. Resonant dielectric elements with high refractive index located on a substrate with low refractive index are adopted in this design. The metasurface is able to induce polarization conversion between reflected and incident waves. By tuning dimensions of the high-refractive-index dielectric elements, phase shift from 0 to 2π can be obtained between reflected and incident waves. Further appropriate adjustment of the dielectric elements dimensions can render the phase difference between different elements largely unchanged in a broad bandwidth from 53 to 70 GHz, offering the ability of producing broadband millimeter wave OAM. Figure of merit such as polarization conversion efficiency and mode spectrum are applied to quantitatively evaluate the quality of the generated OAM beams. Higher-order OAM beam generators are also investigated. Simulation results demonstrate the validity of the proposed structure.

## 2. Design Methodology and Simulation Results
### 2.1 Dielectric Metasurface Design

Metasurfaces are two-dimensional equivalents of metamaterials with thickness much smaller than a wavelength, which is thus endowed with the capability of introducing an abrupt phase change to the field going through or being reflected from it and shaping the wavefront in diversified manner [16]. Dielectric metasurfaces offer a favorable advantage over their metallic counterparts in terms of much less ohmic loss [17].

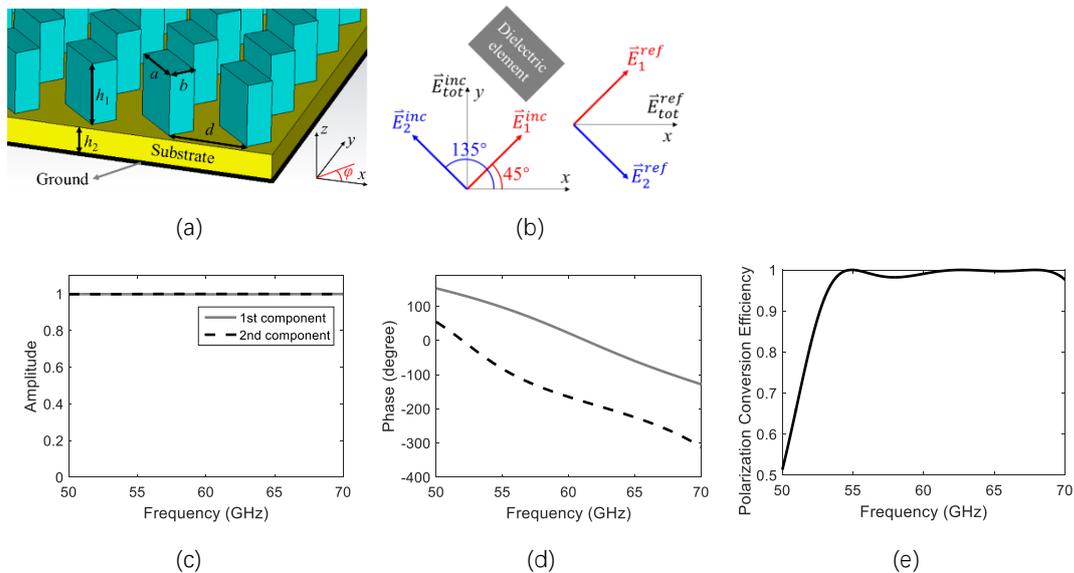

Fig. 1.  (a) Schematic of the dielectric metasurface. (b) Schematic explanation of the polarization conversion effect. Assume the first component keeps the same direction while the second component is reversed in direction due to the 180° phase difference. (c) and (d) are respectively the amplitudes (normalized) and phases of the reflected fields for the two decomposed components in (b). (e) Simulated polarization

conversion efficiency. (c)-(e) are obtained using $a$ = 0.7 mm, $b$ = 0.35 mm, $d$ = 2 mm, $h_1$ = 1.68 mm and $h_2$ = 0.7 mm.

This work employs a dielectric metasurface structure composed of high-refractive-index dielectric elements locating on a low-refractive-index dielectric substrate, as depicted in Fig. 1(a). The substrate is backed by a ground plane, so this dielectric metasurface works in reflective mode. Permittivity of the dielectric elements and substrate are 11.9 and 2.65, respectively, which correspond to silicon and F4B. Both materials are assumed to have negligible loss. The dielectric elements represent good dielectric resonators that can support electric and magnetic dipole responses due to Mie resonances [18].

The high-refractive-index dielectric elements are tilted 45° with respect to the x and y axes, which can induce linear polarization conversion to incident waves polarized along x- or y-direction, i.e., polarization of the reflected wave is rotated by 90° with respect to that of the incident wave. Take a y-direction polarized incident field as an example, the field can be decomposed into two components polarized along $\varphi$ = 45° and 135°, respectively, as shown in Fig. 1(b) as $\vec{E}_1^{inc}$ and $\vec{E}_2^{inc}$. The resultant amplitudes and phases of their reflected waves are plotted in Fig. 1(c) and (d), exhibiting total reflectance and roughly 180° phase difference between the two orthogonal polarizations in the targeted bandwidth from 53 to 70 GHz. Accordingly, the reflected two components $\vec{E}_1^{ref}$ and $\vec{E}_2^{ref}$ effectively form an x-direction polarized reflected field leading to a 90° rotation with respect to the y-direction polarized incident field, shown in Fig. 1(b). Compared with metasurfaces that keep the polarization of the scattered waves, such polarization conversion feature is beneficial to enhancing signal-to-noise ratio for practical applications of metasurfaces [18]. Simulated polarization conversion efficiency, defined as $|E_{cross}|^2/(|E_{cross}|^2 + |E_{co}|^2)$, of the dielectric metasurface is shown in Fig. 1(e). It is apparent that the conversion efficiency maintains over 0.93 from 53 to 70 GHz, which is a key figure of merit of the proposed metasurface.

**2.2 Broadband OAM Generator Design**

An ideal OAM beam bears an azimuthal phase variation represented by $e^{jm\varphi}$. OAM mode number $m$ = −1, with the phase decreasing counterclockwise (along the azimuthal direction) from 0 to $2\pi$ in the xoy plane and the +z axis being propagation direction of the beam, is adopted here as an example to explain the design methodology based on the dielectric metasurface. To realize the azimuthal phase change for $m$ = −1, a metasurface plate is divided into eight sections that are engineered to engender an decremental phase shift of $\pi/4$ to the cross-polarized reflected waves from section 1 to 8, shown in Fig. 2. Accordingly, 0 to −2$\pi$ phase variation can be achieved in a stepped manner along the azimuthal direction. This is implemented by adjusting dimensions of the dielectric elements in the first four sections to obtain, respectively, phase shifts of 0 (section 1 is taken as a phase reference), −$\pi/4$, −$\pi/2$ and −3$\pi/4$ and rotating these four elements by 90° for the rest four sections to attain phase shifts of −$\pi$, −5$\pi/4$, −3$\pi/2$ and −7$\pi/4$. Schematic of the dielectric resonators is shown in Fig. 2.

Generally speaking, the phase shift induced by each section of the metasurface cannot stay the same if frequency is varied, which renders a broadband OAM launcher based on the applied metasurface seemingly impossible. However, maintaining the decremental phase shift to be $\pi/4$ can also effectively accomplish wide operating bandwidth from 53 to 70 GHz for the OAM generator. This is realized by further tuning dimensions of the dielectric resonators via optimization function in CST Microwave Studio. The unit cell period is chosen as $d = 2$ mm, which is subwavelength in the entire design bandwidth. The optimized dimensions of the eight resonators are $a = 0.68$ mm, $b = 0.3$ mm for sections 1 and 5; $a = 0.7$ mm, $b = 0.35$ mm for sections 2 and 6; $a = 0.72$ mm, $b = 0.4$ mm for sections 3 and 7; $a = 0.77$ mm, $b = 0.44$ mm for sections 4 and 8. Optimal thicknesses of the resonators and substrate are found to be $h_1 = 1.68$ mm and $h_2 = 0.7$ mm, respectively.

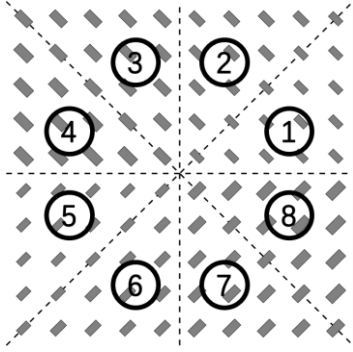

Fig. 2. Schematic of dielectric resonators in the eight sections.

Simulated phase shifts of the cross-polarized reflected waves for the eight resonators are plotted in Fig. 3(a) as a function of frequency. At each simulated frequency, the phase shift of section 1 is chosen as a phase reference and set to 0. It is obvious that the phase differences between each two adjacent sections are approximately $-\pi/4$ in the entire bandwidth, enabling broadband spiral phase profile and associated OAM beam generation. The corresponding amplitude of the cross-polarized reflected waves is also given in Fig. 3(b). Despite that the amplitude drops to 0.78 for resonators 1 and 5 at 53 GHz, all the resonators present an over 82% reflectance across the design bandwidth. Therefore, both the phase shifts and reflected cross-polarized amplitudes of all the designed dielectric resonators satisfy the requirement for a broadband OAM generator.

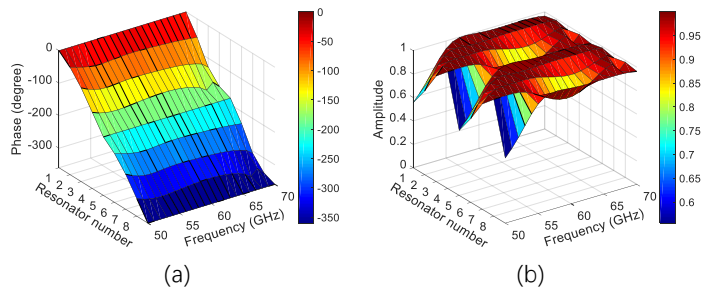

(a)          (b)

Fig. 3. Simulated phase shifts (a) and amplitudes (b) of cross-polarized reflected waves for the eight resonators.

A complete dielectric metasurface plate consisting of 50 × 50 total elements is simulated to visualize the generation of OAM beams. A circular waveguide located on the z axis and 10 cm from the plate serves as a feeding antenna. Simulated phase and amplitude profiles of the cross-polarized reflected electric fields of the created OAM beams 20 cm away from the metasurface plate at 53, 58, 64 and 70 GHz are provided in Fig. 4, which unambiguously manifest the spiral phase nature and doughnut-shaped amplitude pattern of OAM beams. To quantitatively evaluate quality of the OAM beams, OAM mode spectra [8] are also calculated and shown in Fig. 4, from which it is seen that the mode $m = -1$ is predominant at 53, 58, 64 and 70 GHz. This is also the case for the OAM beams at other frequencies in the design frequency band.

OAM generators with higher mode numbers $m = -2$ and $-3$ are also designed with the identical eight dielectric resonators used for the $m = -1$ generator. The former metasurface is divided into 16 sections and the latter one incorporates 24 sections, still maintaining an decremental phase shift of $-\pi/4$ between each two adjacent sections. Simulated phase and amplitude profiles of the created high-order OAM beams as well as their mode spectra are given in Fig. 5 and 6. The phase profiles exhibit correct spiral patterns for both the higher-order OAM modes at 53, 58, 64 and 70 GHz. The amplitude profiles are still approximately donut-shaped. The corresponding mode spectra also demonstrate that the $m = -2$ and $-3$ modes dominate in the spectra at all the simulated frequencies. Therefore, the proposed metamaterial structure is valid for launching higher-order OAM modes.

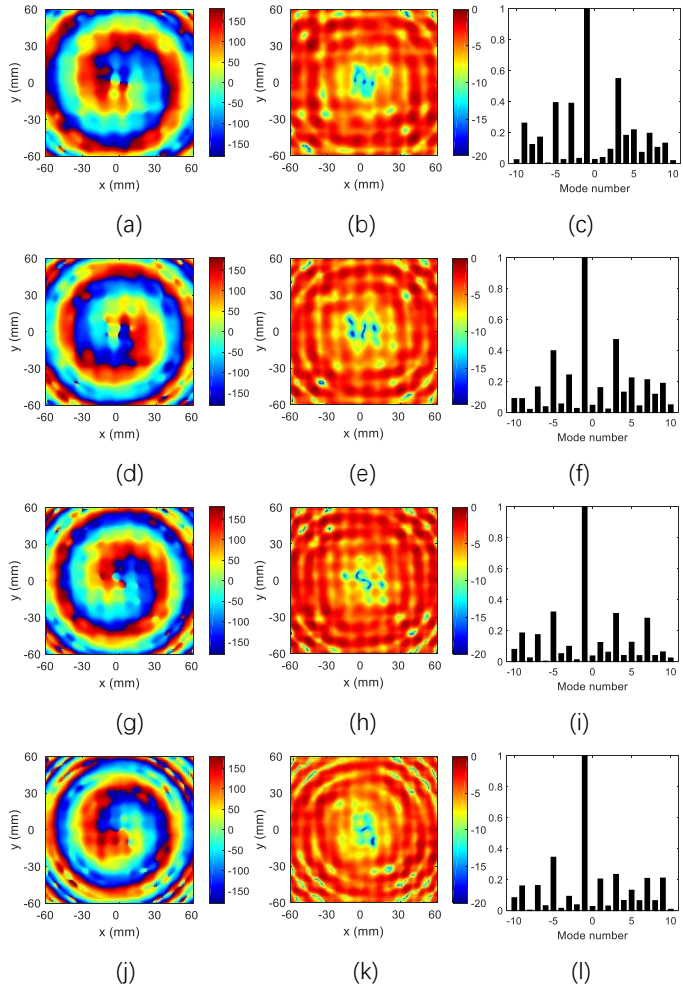

Fig. 4.  Simulated phase profiles (in degree), normalized amplitude profiles (in dB scale) and mode spectra

of the cross-polarized electric fields of the generated $m = -1$ OAM beam at 53 GHz (a)-(c), 58 GHz (d)-(f), 64 GHz (g)-(i), and 70 GHz (j)-(l).

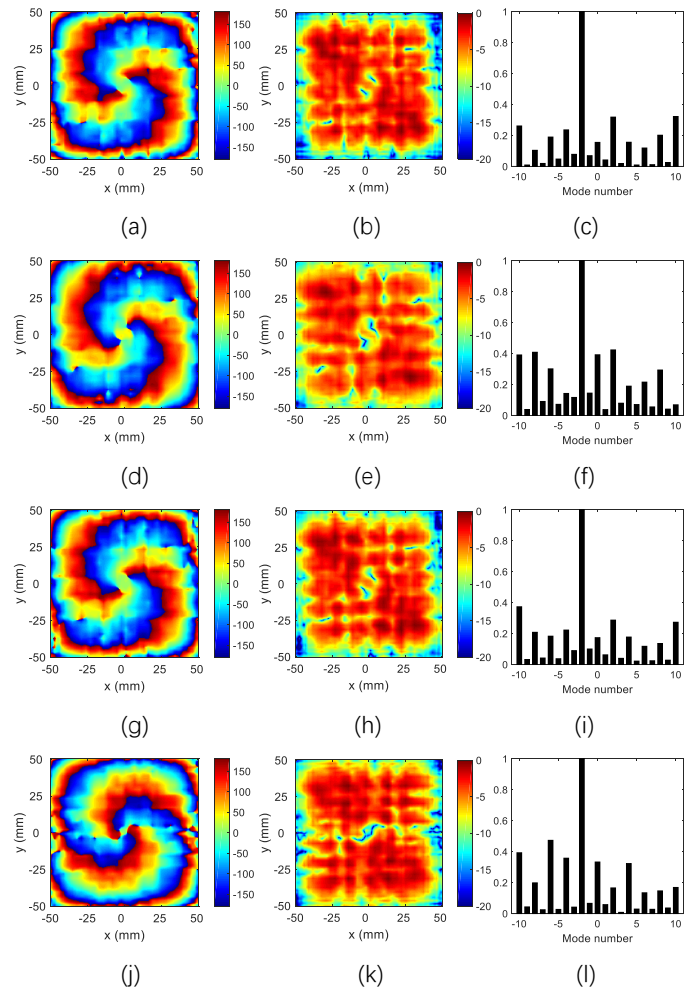

Fig. 5. Simulated phase profiles (in degree), normalized amplitude profiles (in dB scale) and mode spectra of the cross-polarized electric fields of the generated $m = -2$ OAM beam at 53 GHz (a)-(c), 58 GHz (d)-(f), 64 GHz (g)-(i), and 70 GHz (j)-(l).

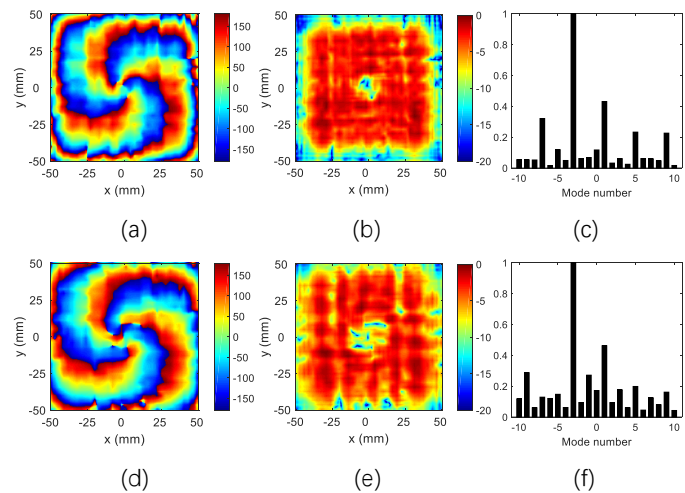

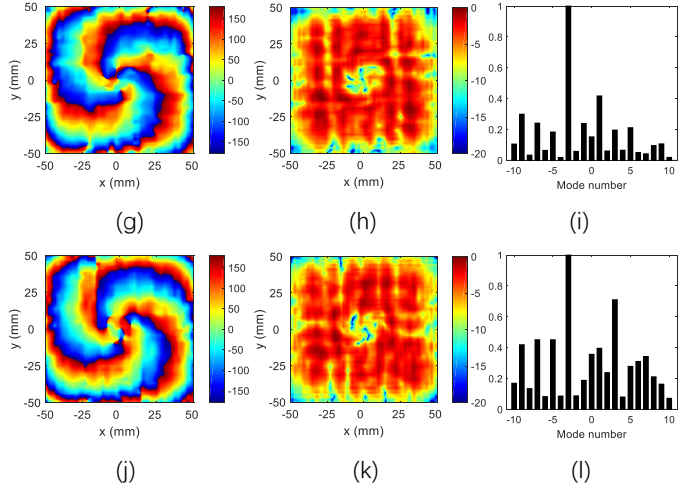

Fig. 6.  Simulated phase profiles (in degree), normalized amplitude profiles (in dB scale) and mode spectra of the cross-polarized electric fields of the generated $m = -3$ OAM beam at 53 GHz (a)-(c), 58 GHz (d)-(f), 64 GHz (g)-(i), and 70 GHz (j)-(l).

## 3. Conclusions

A millimeter wave broadband OAM generator using reflection-type dielectric metasurface is proposed in this work. Detailed design method and working principle of the structure is introduced. High quality of the generated OAM beams bearing distinct mode numbers are demonstrated by simulated phase profiles, amplitude patterns and mode spectra.

**Acknowledgments**

This work was supported in part by the National Nature Science Foundation of China under grant 61701305 and the Shanghai Pujiang Program under grant 17PJ1406600.